\documentclass[conference]{IEEEtran}
\IEEEoverridecommandlockouts
\usepackage{tikz}
\usepackage[siunitx]{circuitikz}
\usepackage{cite}
\usepackage{amsmath,amssymb,amsfonts}
\usepackage{algorithmic}
\usepackage{graphicx}
\usepackage{textcomp}
\usepackage{xcolor}
\usepackage{multirow}
\usepackage{booktabs,tabularx}
\usepackage{bm}
\usepackage{enumitem}
\usepackage{subfiles}
\usepackage{amsthm}
\usepackage{dblfloatfix}
\usepackage{float}
\usepackage{hyperref}
\usepackage{oplotsymbl}
\usepackage{amsmath}
\usepackage[font=footnotesize]{caption}
\usepackage[font=footnotesize]{subcaption}
\usepackage{subcaption}
\usepackage{epstopdf}

\newcommand{\normtwo}[1]{\left\lVert#1\right\rVert_2}
\newcommand*\xbar[1]{%
  \hbox{%
    \vbox{%
      \hrule height 0.5pt 
      \kern0.2ex
      \hbox{%
        \kern-0.1em
        \ensuremath{#1}%
        \kern-0.1em
      }%
    }%
  }%
}

\def\BibTeX{{\rm B\kern-.05em{\sc i\kern-.025em b}\kern-.08em
    T\kern-.1667em\lower.7ex\hbox{E}\kern-.125emX}}

\begin{document}
\title{Continual Learning-Based MIMO Channel Estimation: A Benchmarking Study\\
\thanks{This work was partially supported by the Discovery Grants Program of the Natural Sciences and Engineering Research Council of Canada (NSERC).}
}

\author{\IEEEauthorblockN{Mohamed Akrout$^*$, Amal Feriani$^*$, Faouzi Bellili, Amine Mezghani, and Ekram Hossain}
\IEEEauthorblockA{Department of Electrical and Computer Engineering, University of Manitoba, Winnipeg, Canada \\
\{akroutm,\,feriania\}@myumanitoba.ca, \{Faouzi.Bellili,\,Amine.Mezghani,\,Ekram.Hossain\}@umanitoba.ca}
}

\maketitle

\def\thefootnote{*}\footnotetext{equal contribution, order determined randomly.}

\thispagestyle{empty}
\pagestyle{empty}

\begin{abstract}
With the proliferation of deep learning techniques for wireless communication, several works have adopted learning-based approaches to solve the channel estimation problem. While these methods are usually promoted for their computational efficiency at inference time, their use is restricted to specific stationary training settings in terms of communication system parameters, e.g., signal-to-noise ratio (SNR) and coherence time. Therefore, the performance of these learning-based solutions will degrade when the models are tested on different settings than the ones used for training. This motivates our work in which we investigate continual supervised learning (CL) to mitigate the shortcomings of the current approaches. In particular, we design a set of channel estimation tasks wherein we vary different parameters of the channel model. We focus on Gauss-Markov Rayleigh fading channel estimation to assess the impact of non-stationarity on performance in terms of the mean square error (MSE) criterion. We study a selection of state-of-the-art CL methods and we showcase empirically the importance of catastrophic forgetting in continuously evolving channel settings. Our results demonstrate that the CL algorithms can improve the interference performance in two channel estimation tasks governed by changes in the SNR level and coherence time.


\end{abstract}


\section{Introduction}
\subsection{Background and motivation}

Deep neural networks (DNN) are widely applied to solve wireless communication problems, such as the channel estimation problem, where they played important roles
in improving the system performance and reducing computational complexity \cite{le2021deep,hu2020deep}. However, these learning-based methods are usually trained and evaluated on \emph{fixed datasets} and/or \emph{stationary propagation environments} \cite{shen2019lorm}. Indeed, critical parameters of communication system models such as the signal-to-noise ratio (SNR) and coherence time ($T_c$) are often assumed to be fixed for these learning-based methods. Since wireless communications systems are \emph{inherently non-stationary}, trained models need to be tested on continuously and dynamically changing datasets. This shift in data distribution results in the degradation or even the failure of the trained DNNs, and this problem is known in the literature as \emph{catastrophic forgetting} \cite{french1999catastrophic}. One solution to avoid catastrophic forgetting is to train an ensemble or a pool of models for different training scenarios (a.k.a., tasks) and use the appropriate model at inference time. Such approaches are inefficient, because $i)$ they do not scale well since the number of models will increase as a function of the number of tasks, and $ii)$ the appropriate model selection at inference time is challenging. Thus, there is an urge to design solutions that can learn incrementally and adapt to changes in the environment.

Continual learning (CL) methods \cite{DBLP:journals/pami/LangeAMPJLST22, hadsell2020embracing, DBLP:journals/corr/abs-1904-07734} were introduced to train DNNs on sequentially changing tasks without forgetting the previously learned ones. CL training strategies sidestep the impact of averaging conflicting gradients on the learning dynamics of DNNs \cite{hadsell2020embracing}. As one example, consider training a DNN sequentially on two different tasks $\mathcal{T}_{\textrm{SNR,high}}$ then $\mathcal{T}_{\textrm{SNR,low}}$ at high and low SNR levels, respectively. Once the training on the first task $\mathcal{T}_{\textrm{SNR,high}}$ is done, an equilibrium in the space of DNN parameters is reached ensuring good performance on the task $\mathcal{T}_{\textrm{SNR,high}}$. The subsequent training on the second task  $\mathcal{T}_{\textrm{SNR,low}}$, however, will pull all the DNN parameters toward the direction of its gradients, thereby leading to forgetting the first task $\mathcal{T}_{\textrm{SNR,high}}$. For this reason, gradient-based training algorithms require that, in expectation, all tasks are present to keep the conflicting tension between gradients stemming from multiple tasks \cite{hadsell2020embracing}. With CL techniques, the DNN is expected to adapt to task changes by not forgetting the learned skill used in the previous task. 

There are different CL scenarios depending on the level of difficulty \cite{DBLP:journals/corr/abs-1904-07734}. The easiest one, called \textbf{task-incremental learning (Task-IL)}, assumes the identity of the task to be always provided which makes task-specific training possible. The typical approach for this scenario is to train a ``multi-headed” DNN where each task has its output units and the rest of the network is shared between tasks. The second scenario is \textbf{domain-incremental learning (Domain-IL)} where the task identity is not known at inference time. A common example in this scenario is where the input distribution is changing but the nature of the task remains the same.

In this paper, we formulate the wireless channel estimation problem in a multiple-input multiple-output (MIMO) wireless system as a domain-IL problem for two main reasons: $i)$ the distribution shift in the DNN inputs due to changes in the propagation environment properties, and $ii)$ the DNN is expected to seamlessly estimate the channel, i.e., without knowing when the environment has changed.

\subsection{Prior Work}
\subsubsection{Overview of continual learning methods}\label{subsubsec:overview-cl}
Continual learning is an active research field within the machine learning community \cite{dohare2021continual}. To tackle catastrophic forgetting, CL-based methods can be divided into three categories  based on how task information is stored and used throughout the sequential learning process (see \cite{DBLP:journals/pami/LangeAMPJLST22} for a comprehensive review):

\noindent \textbf{Replay methods.}~These methods store a subset of the training task samples in a memory buffer and revisit them while learning a new task. The task samples are either reused as DNN inputs for rehearsal \cite{rebuffi2017icarl,rolnick2019experience,isele2018selective}, or to constrain the optimization of the new task loss to prevent previous task interference \cite{lopez2017gradient,chaudhry2018efficient}.

\noindent \textbf{Regularization-based methods.}~Unlike replay methods, regularization-based methods do not store task samples to respect the data privacy, which also eases the memory requirements \cite{li2017learning,kirkpatrick2017overcoming,nguyen2017variational}. Alternatively, they introduce an extra regularization term in the loss function of the DNN, thereby consolidating previous knowledge when learning on new tasks.

\noindent \textbf{Parameter isolation methods.}~This family of methods append new model parameters to each task. Therefore, DNNs can grow new layer branches dedicated for the training for new tasks, while freezing \cite{xu2018reinforced} or masking out \cite{mallya2018packnet,serra2018overcoming} branches trained on previous tasks.

\noindent For real-world problems, the aforementioned CL categories should satisfy the following three constraints \cite{DBLP:journals/pami/LangeAMPJLST22}:
\begin{itemize}[leftmargin=*]
    \item \textit{No task boundaries}: tasks do not arrive sequentially and are not delineated by clear boundaries during training,
    \item \textit{No task identifiers}: the inference step does not identify the task difficulty to the DNN using task identifiers,
    \item \textit{Constant memory}: the task dataset collected during the training should fit a buffer with a bounded memory footprint.\vspace{0.2cm}
\end{itemize}

\subsubsection{Continual learning for wireless communications}
Little effort has been devoted to investigate CL methods in wireless communications problems. A CL formulation for wireless system design was adopted in \cite{sun2022learning} to optimize wireless resource allocation in a dynamic environment. In this work, the DNN makes use of a continuously updated data buffer containing representative historical data from different dynamic settings of the environment. Another line of work proposed a CL-based minimum mean-square error (CL-MMSE) technique for mmWave channel estimation to address high frequency signal characteristics \cite{kumar2021continual}. However, CL-MMSE ignores pilot transmission overhead and only varies the number of receive antennas between 8 and 128 to generate the CL tasks.

 

\subsection{Contributions}
Different from the aforementioned work, we present an exhaustive benchmark of CL approaches for MIMO channel estimation.
The contributions of this paper are as follows:
\begin{itemize}[leftmargin=*]
    \item We formulate the channel estimation problem as a domain-IL problem and propose two different benchmarks based on the coherence time and SNR considerations;
    \item We conduct an exhaustive empirical comparison of seven CL methods with a detailed analysis of their performance;
    \item For the first time, we provide a  benchmark for CL for wireless communications.
\end{itemize}

\section{System Model and Assumptions}\label{sec:sm}

\begin{figure}[h!]
    \centering
    \includegraphics[scale=0.4]{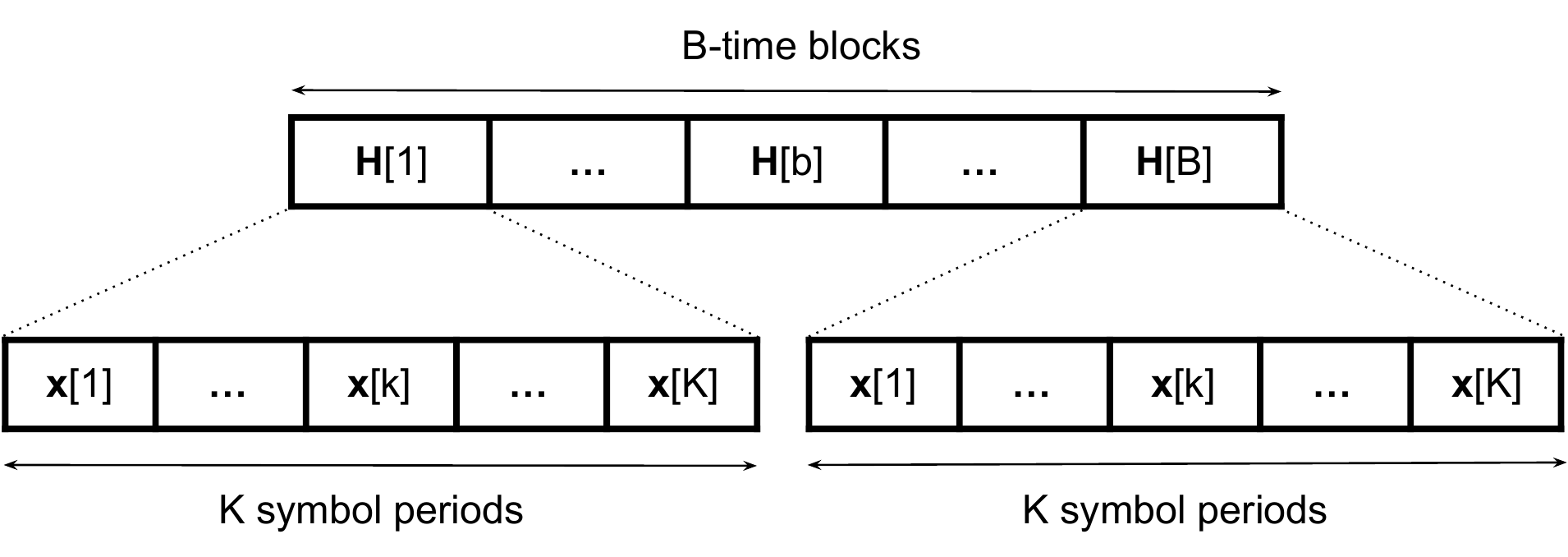}
    \caption{ An illustration of the considered channel estimation method. The training time is divided into $B$ time-blocks of duration $K$ symbol periods each. The channel is constant during the $b$-th time-block and varies between different time-blocks while the same pilot symbols $\mathbf{x}[1]$, . . . , $\mathbf{x}[K]$ are reused from one block to another.}
    \label{fig:time-slots}
\end{figure}

We consider a MIMO communication system under Gauss-Markov Rayleigh fading channels between $N$ transmit antennas and $M$ receive antennas. The transmission lasts for $B$ time blocks each of which is used to transmit $K$ symbols. The channel estimation method is illustrated in Fig.~\ref{fig:time-slots} where the block-fading channel is assumed to remain constant over the entire $b$-th time block ($b\in\{1,\dots,B\}$), while the total duration of $KB$ symbol periods is used for channel estimation within the overestimated channel coherence time $T_c$.\vspace{-0.25cm}

\noindent For every time-block $b$, the received signal is given by:

\begin{equation}\label{eq:received-signal}
    \bm{y}_b = \bm{H}_b \,\bm{x}_b + \bm{n}_b,\,\quad\quad b=1,\dots,B.
\end{equation}
\noindent In (\ref{eq:received-signal}), $\bm{x}_b\in \mathbb{C}^{N}$ represents the transmitted symbol vector and $\bm{n}_b \in \mathbb{C}^{M}$ is an additive complex white Gaussian noise vector, i.e., $\bm{n}_b \sim \mathcal{CN}(\mathbf{0}_{M},\sigma^2_{\bm{n}_b}\,\mathbf{I}_{M})$. Here, at a given SNR level $\rho$ [dB], the noise variance $\sigma^2_{\bm{n}_b}$ is defined as:
\begin{equation}\label{eq:noise-variance}
    \sigma^2_{\bm{n}_b} = \mathbb{E}\Big[\normtwo{\bm{H}_b \,\bm{x}_b}^2\Big]\,10^{-\rho/10}.
\end{equation}

\noindent Moreover, $\bm{H}_b$ is the time-varying Rayleigh fading channel evolving according to the Gauss-Markov model \cite{jakes1994microwave}:
\begin{equation}\label{eq:gauss-markov-evolution}
    \bm{H}_b = \sqrt{\alpha}\,\bm{H}_{b-1} + \sqrt{1-\alpha}\,\bm{W}_{b},\,\quad\quad b=2,\dots,B.
\end{equation}
\noindent where $\alpha\in [0,1[$ is the channel memory factor. Moreover, $\bm{H}_1$ and $\bm{W}_b$ are independent and identically distributed realizations drawn from the complex Gaussian distribution $\mathcal{CN}(\mathbf{0}_{NM},\mathbf{I}_{NM})$. By recursively injecting the expression of $\bm{H}_{b-1}$ in (\ref{eq:gauss-markov-evolution}) for $b=1,\dots,B$, we obtain:
\begin{equation}\label{eq:non-recursive-gauss-markov-evolution}
    \bm{H}_b = \sqrt{\alpha}^{b-1}\,\bm{H}_{1} + \sqrt{1-\alpha}\,\sum\limits_{p=2}^{b}\,\sqrt{\alpha}^{b-p}\,\bm{W}_{p}.
\end{equation}
\noindent Therefore, the sequence $\{\bm{H}_b\}_b$ in (\ref{eq:non-recursive-gauss-markov-evolution}) forms a discrete-time Markov chain that evolves as a random walk process starting from the initial channel $\bm{H}_1$ scaled by the $b$-th block-dependent factor $\sqrt{\alpha}^{b-1}$. This Markov chain  models the stochastic channel aging which induces significant shifts in channel distribution. The channel aging phenomenon results from the variation of the channel between the instant when it is learned using pilots and the instant when it is used for signal processing. This time variation is due to user mobility and processing delay.



\begin{figure*}[!th]
     \centering
     \begin{subfigure}[b]{0.4\textwidth}
         \centering
         \includegraphics[width=\textwidth]{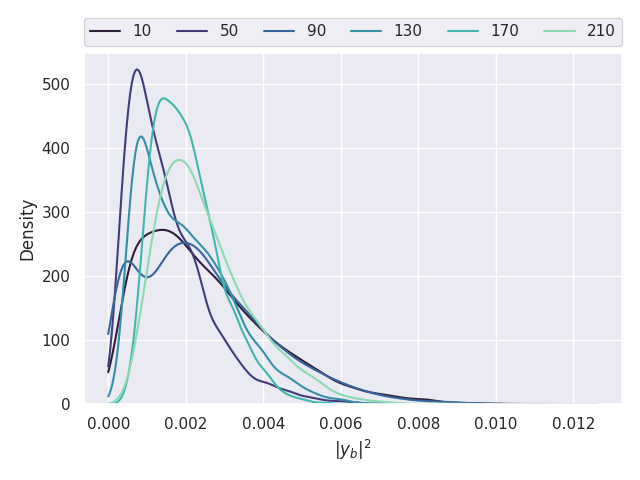}
         \caption{varying $T_c$ at $\rho=10\, \textrm{dB}$}
         \label{fig:dist-shift-tc}
     \end{subfigure}
     \begin{subfigure}[b]{0.4\textwidth}
         \centering
         \includegraphics[width=\textwidth]{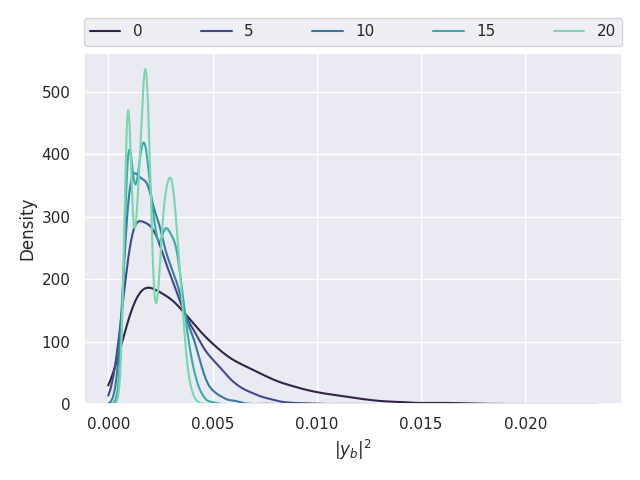}
         \caption{varying $\rho$ at $T_c= 20~\textrm{symbol periods}$}
         \label{fig:dist-shift-snr}
     \end{subfigure}
    \caption{Illustration of the distribution shift in the received signal due to changes in (a) the time coherence $T_c$ and (b) the SNR level $\rho$.}
    \label{fig:dist-shift-example}
    \vspace{-0.3cm}
\end{figure*}

\section{Continual Learning-Based Channel Estimation}\label{sec:cl}
\subsection{Motivating Example}\label{subsec:motivating-example}
As mentioned before, the wireless channel is non-stationary, thus in this example, we will study the impact of the non-stationarity of the channel $\bm{H}_b$ on the distribution of the received signal $\bm{y}_b$ given in (\ref{eq:received-signal}). This is because $\bm{y}_b$ is the input vector to the neural network for channel estimation. In this context, we consider two main changing factors impacting the received signal $\bm{y}_b$: $i)$ the SNR level $\rho$ in (\ref{eq:noise-variance}) and the coherence time $T_c = K$ as depicted in Fig.~\ref{fig:time-slots}.

For better illustration, we consider the MIMO communication system described in Section \ref{sec:sm} with $N=8$ and $M=5$. We then examine the following two scenarios:

\begin{itemize}[leftmargin=*]
    \item In Fig. \ref{fig:dist-shift-tc}, we fix the SNR level to $\rho=10$ dB and vary the coherence time $T_c$ between $10$ and $250$ transmit symbol periods. We observe a positive shift (i.e., translation) in the empirical density of the received signal as the coherence time increases\footnote{\label{density-comment}Here the density has higher values than 1 which does not contradict basic probability theory because the density corresponds to the probability per unit value \big(here $\sim 10^{-3}$\big) of the random variable representing the received power.}.
    \item In Fig. \ref{fig:dist-shift-snr}, we fix the coherence time $T_c$ to $20$ symbol periods and vary the SNR between $0$ and $25$ dB. It is seen that the empirical density of the received power\textsuperscript{\ref{density-comment}} is unimodal at lower SNR levels (i.e., 0, 5 and 10 dB) and becomes multimodal at higher SNR levels (i.e., 15 and 20 dB).
\end{itemize}

This example showcases that the change in the characteristics of the wireless channel will lead to a shift in the distribution of the received signal. This motivates the design of channel estimation algorithms based on CL to avoid catastrophic forgetting resulting from the non-stationarity of wireless channels. In the next section, we explain the design of these CL tasks for channel estimation.



\subsection{Task protocols}\label{subsec:task-protocol}

To comprehensively compare the performance of various CL approaches for wireless channel estimation, we consider two sets of domain-IL tasks, $\mathcal{T}_{\textrm{SNR}}$ and $\mathcal{T}_{T_c}$, defined as follows:\vspace{0.2cm}
\begin{itemize}[leftmargin=*]
    \item  $\mathcal{T}_{\textrm{SNR}} \triangleq \bigcup\limits_{i}^{}\big\{\mathcal{T}_{\textrm{SNR}=i [\textrm{dB}]}\big\}$ where different channel estimation tasks are generated by varying the SNR level $\rho$. In this work, we assume the SNR ranges from $0\,\textrm{dB}$ to $20\,\textrm{dB}$ with a step of $2$ dB, resulting in $11$ tasks in total;
    \item $\mathcal{T}_{T_c} \triangleq \bigcup\limits_{j}^{}\big\{\mathcal{T}_{T_c=j}\big\}$ where a sequence of $20$ tasks are generated by varying the index $j$ between $5$ and $100$ symbol periods with a step size of $5$ symbol periods.
\end{itemize}

\noindent Note that we consider these tasks as domain-IL since an \emph{implicit} task-dependent transformation will be applied on the inputs of the channel estimation model (i.e., received signal). However, it is important to highlight that this setting is slightly different from the common assumption in the CL literature where domain-IL tasks are defined using \emph{direct} transformation on the inputs (e.g., rotation, permutation). 

\noindent Furthermore, we scrutinize how changing the task
order affects the channel estimation. We do so by considering two different task orderings of the sets $\mathcal{T}_{\textrm{SNR}}$ and $\mathcal{T}_{T_c}$:
\begin{itemize}[leftmargin=*]
    \item \textbf{Curriculum}: the tasks are ordered from easy to hard which
imitates the meaningful learning order in human curricula. For instance, tasks are presented to the DNN from high to low SNR and $T_c$ values;
    \item \textbf{Random}: the tasks are presented to the DNN in a random order.
\end{itemize}
\noindent The task ordering hypothesis suggests that starting from the curriculum order of tasks would lead to a better result as compared to the random order of tasks, and some CL methods have shown little sensitivity to the task order in the context of image classification \cite{DBLP:journals/pami/LangeAMPJLST22}. Whether the same behavior is observed for channel estimation problems will be discussed in Section \ref{sec:results}.

\section{Experimental Setup}

\subsection{Benchmarking Methods}\label{subsec:benchmark-algorithms}
We consider the following CL methods:
\begin{itemize}[leftmargin=*]
    \item \textbf{Synaptic Intelligence (SI)} \cite{zenke2017SI}: this is a regularization-based method in which a regularization term is added to penalize changes in parameters when they move away from their values after training on previous tasks; 
    \item \textbf{Experience Replay (ER)} \cite{rebuffi2017icarl,riemer2018MER,robins1995catastrophic}: this replay-based method concatenates the training samples from the current task with old samples from previous tasks to construct training batches; 
    \item  \textbf{Averaged-GEM (A-GEM)} \cite{chaudhry2018AGEM}: this is an lightweight version of the Gradient Episodic Memory (GEM) \cite{lopez2017GEM} that uses an episodic memory to leverage old training data and introduce constraints to maintain similar predictions on previous tasks while learning the new one. This method uses task boundaries to store the task samples in the buffer;
    \item \textbf{Averaged-GEM with a reservoir buffer (A-GEM-R)} \cite{buzzega2020DER}: this method is similar to A-GEM. The main difference is in how the replay buffer is populated. This method uses reservoir memory buffer to avoid using task boundaries;
    \item \textbf{Function Distance Regularization (FDR)} \cite{benjamin2018FDR}: this is another replay-based method in which the network outputs and training samples are stored to align previous and current outputs using an $L_2$-regularization loss. FDR relies on task boundaries to populate the replay buffer;
    \item \textbf{Dark Experience Replay (DER)} \cite{buzzega2020DER}: this is also a replay-based method in which the outputs of the model's final layer are stored instead of the ground truth for the previously learned tasks. This method adopts reservoir sampling to avoid using task boundaries. A penalty term is added to the loss to minimize the difference between the current and the previous model outputs;  
    \item \textbf{Dark Experience Replay++ (DER++)} \cite{buzzega2020DER}: this method is an extension of the DER algorithm where, in addition to the model outputs, the ground truth of previous tasks is stored and another penalty term is added to reduce the difference between the output of the model on the buffer datapoints and their ground truth labels.
\end{itemize}
We further include the following two baselines: 
\begin{itemize}[leftmargin=*]
    \item \textbf{No-CL}: the DNN is trained sequentially on all the tasks without taking into consideration the catastrophic forgetting. This can be seen as an upper bound on the MSE;
    \item \textbf{\textbf{Joint training}}: the DNN is trained on the joint data from all the tasks, which leads to a lower bound on the MSE.
\end{itemize}

\noindent The implementation of the different CL methods is based on the one published in \cite{DBLP:conf/nips/BuzzegaBPAC20}. For a broader reproducible benchmark of CL methods in wireless communication, we will publish on Github the SNR and $T_c$ tasks as well as instructions needed to reproduce the main experimental results presented in Section \ref{sec:results}.
\subsection{Training Details}
\subsubsection{Hyperparameters}
We simulate a MIMO system as described in Section \ref{sec:sm} with $N=8$ and $M=5$. We set the carrier frequency $f_c$ to $3$ GHz, the path-loss exponent to $2.5$, the channel memory factor $\alpha$ to $0.01$. and the distance between the transmitter and the receiver to $20$ meters. We also consider a 2-PAM pilot sequence of $10$ symbols.

\subsubsection{Dataset generation}
A dataset of $20000$ samples is generated following the system dynamics described in Section \ref{sec:sm}. The generated data is shuffled and divided such that $90\%$ of the generated data is used for training and the remaining is reserved for testing. 

\subsubsection{Architecture and training}
For all the tasks, we use a fully-connected neural network with five hidden layers of $3\times M$ ReLU units followed by an output layer with $N$ units. All models are trained using stochastic gradient descent for a fair comparison between all CL methods. 
For each task, the model is trained for $100$ epochs to reach a satisfactory performance. Due to computational complexity, we avoid the online scenario (i.e., train the model for one epoch) since it will not be possible to distinguish between the effects of forgetting and under-fitting \cite{DBLP:conf/nips/BuzzegaBPAC20}. 

\subsection{Evaluation Metrics}

While monitoring the MSE performance of the model across tasks, we assess its ability to transfer knowledge by measuring:
\begin{itemize}[leftmargin=*]
    \item \textit{The backward transfer} which quantifies the influence of learning a task $t_i$ on the performance of a \textit{previous} task $t_k$ for $k<i$. A positive (resp. negative) backward transfer is observed when learning a particular task $t_i$ increases (resp. decreases) the performance on some preceding task $t_k$. Large negative backward transfer is also known as catastrophic forgetting;
    \item \textit{The forward transfer} which describes the impact of learning a task $t_i$ on the performance of \text{future} tasks $t_k$ for $k>i$. Positive forward transfer is observed when the model is able to perform ``zero-shot'' learning by exploiting the feature correlation across tasks.
\end{itemize}

\noindent These two transfer measures are the common evaluation metrics adopted in the CL literature \cite{DBLP:conf/nips/Lopez-PazR17,DBLP:conf/nips/BuzzegaBPAC20,DBLP:journals/pami/LangeAMPJLST22}. More specifically, the model is evaluated on all previously learned tasks after it has finished learning a task $t_i$. Hence, after learning all the $T$ tasks, we get an error matrix $E \in \mathbb{R}^{T\times T}$, where each element $e_{i,j}$ is the MSE error of the task $t_j$ after learning the task $t_i$ (e.g., $e_{T,i}$ stands for the error on the task $i$ after learning all the $T$ tasks). Further, we denote by $\Bar{b}$ the error vector of a randomly-initialized model. The considered evaluation metrics are defined as follows \cite{DBLP:conf/nips/BuzzegaBPAC20}:
\begin{subequations}\label{eq:metrics}
\fontsize{9.5}{9.5}
\begin{align}
    \textbf{Average MSE}~(\texttt{AVG\_MSE}) &= \frac{1}{T} \sum_{i=1}^{T} e_{T,i},\\
    \textbf{Forward transfer}~(\texttt{FWT}) &= \frac{1}{T-1} \sum_{i=2}^{T} \left(\Bar{b}_i- e_{i-1,i}\right),\\
    \textbf{Backward transfer}~ (\texttt{BWT}) &= \frac{1}{T-1} \sum_{i=1}^{T-1} \left(e_{T,i} - e_{i,i}\right).
\end{align}
\end{subequations}


\section{Numerical Results and Analysis}\label{sec:results}
\begin{figure*}[!bh]
\centering
\subcaptionbox{$\mathcal{T}_{\textrm{SNR}}$\,--\,Curriculum}[.245\linewidth][c]{%
    \includegraphics[width=\linewidth]{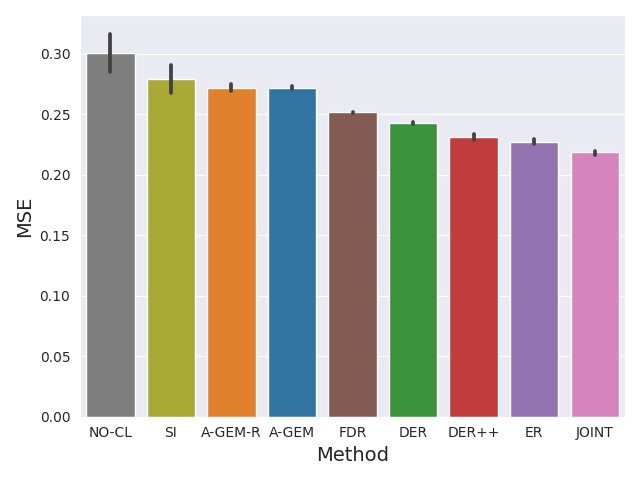}}
\subcaptionbox{$\mathcal{T}_{\textrm{SNR}}$\,--\,Random}[.245\linewidth][c]{%
    \includegraphics[width=\linewidth]{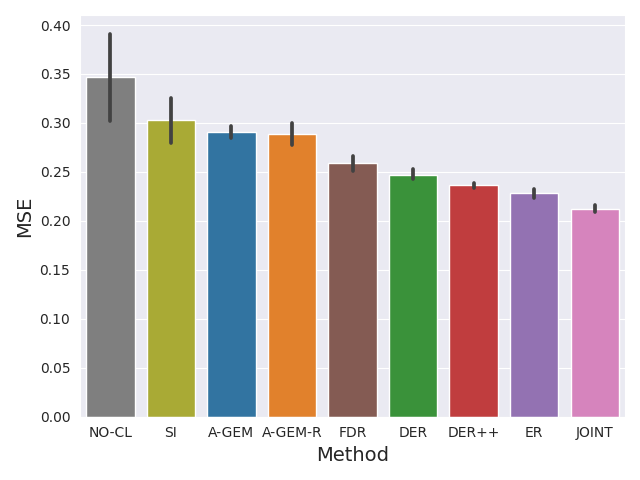}}
\subcaptionbox{$\mathcal{T}_{T_c}$\,--\,Curriculum}[.245\linewidth][c]{%
    \includegraphics[width=\linewidth]{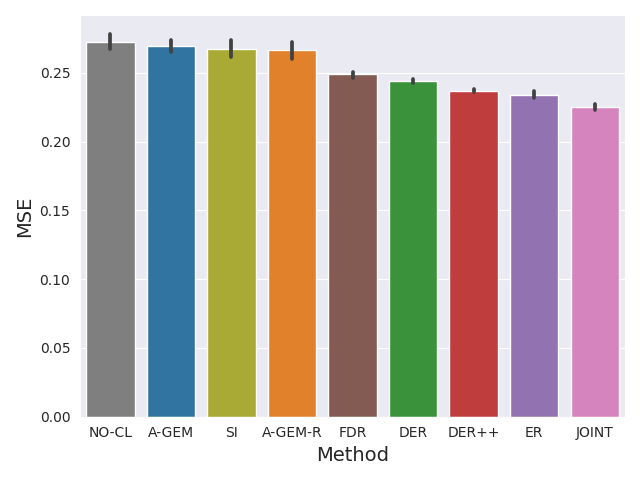}}
\subcaptionbox{$\mathcal{T}_{T_c}$\,--\,Random}[.245\linewidth][c]{%
    \includegraphics[width=\linewidth]{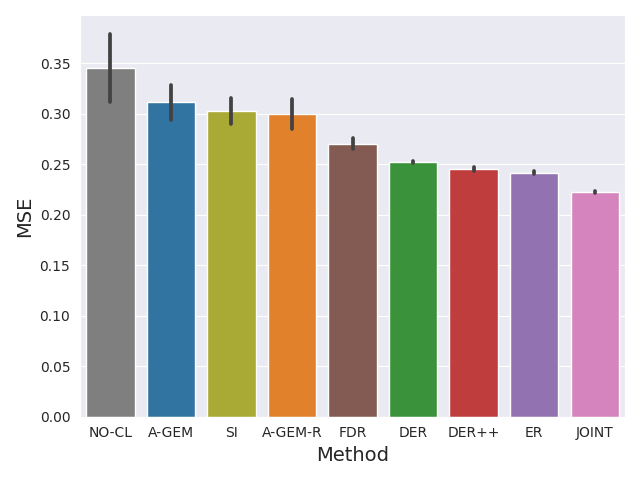}}
\bigskip \vfill
\subcaptionbox{$\mathcal{T}_{\textrm{SNR}}$\,--\,Curriculum}[.24\linewidth][c]{%
    \includegraphics[width=\linewidth]{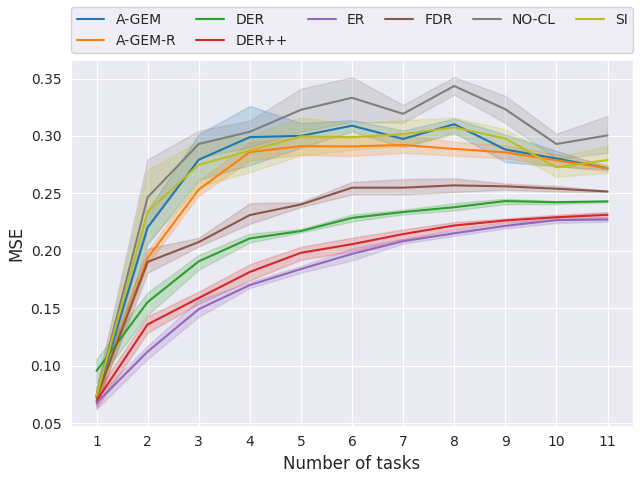}}
\subcaptionbox{$\mathcal{T}_{\textrm{SNR}}$\,--\,Random}[.24\linewidth][c]{%
    \includegraphics[width=\linewidth]{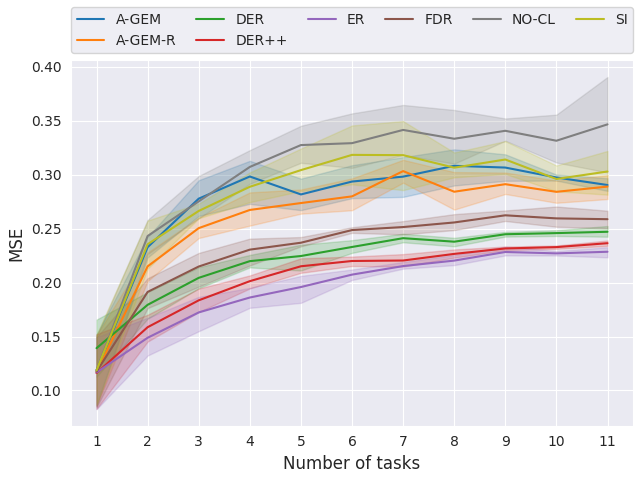}}
\subcaptionbox{$\mathcal{T}_{T_c}$\,--\,Curriculum}[.24\linewidth][c]{%
    \includegraphics[width=\linewidth]{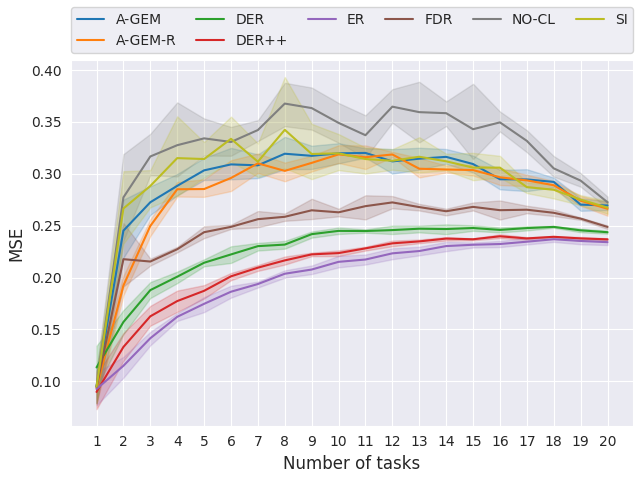}}
\subcaptionbox{$\mathcal{T}_{T_c}$\,--\,Random}[.24\linewidth][c]{%
    \includegraphics[width=\linewidth]{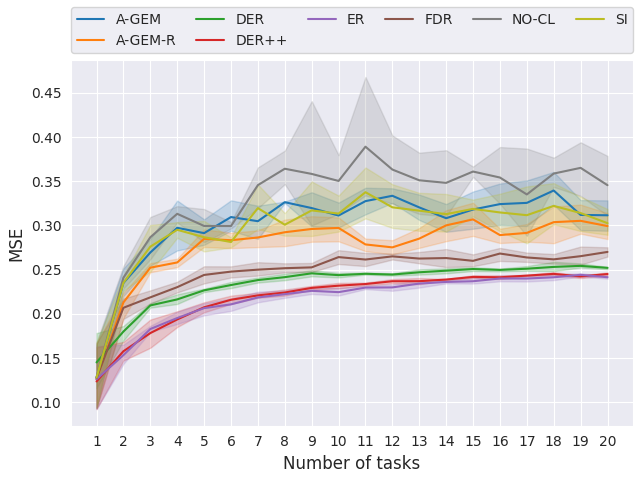}}
\caption{Top: The final MSE after learning all the tasks; Bottom: the evolution of the MSE as more tasks are learned}\label{fig:mse}
\rule{\textwidth}{0.4pt}\vspace{0.2cm}
\end{figure*}

\begin{table*}[!bh]
\caption{Forward and backward transfer for all tasks}
\centering
\begin{tabular}{clcccccccc}
\toprule
    \multirow{2}{*}{Task} & \multirow{2}{*}{Metric} &\multicolumn{8}{c}{CL Methods}\\
    \cmidrule[1pt]{3-10}

    &  &\textbf{NO-CL}& \textbf{SI}& \textbf{A-GEM} & \textbf{A-GEM-R} & \textbf{ER} & \textbf{FDR} & \textbf{DER} & \textbf{DER++}\\
    \midrule
        \multirow{2}{*}{$\mathcal{T}_{\text{SNR}}$\,--\,Curriculum} & \texttt{FWT} $\Uparrow$ & 0.14 & 0.16 & 0.15 & 0.14  & 0.16 & 0.18 & \textbf{0.2} & \textbf{0.2} \\
    \cmidrule(lr{1em}){3-10}
    & \texttt{BWT} $\Downarrow$ & 0.2 & 0.15 & 0.15 & 0.16 & 0.1 & 0.12 & 0.09 & \textbf{0.07} \\
    \midrule
    \multirow{2}{*}{$\mathcal{T}_{\text{SNR}}$\,--\,Random} & \texttt{FWT} $\Uparrow$ & 0.12 & 0.15 & 0.14 & 0.14  & 0.16 & 0.18 & \textbf{0.2} & 0.19 \\
    \cmidrule(lr{1em}){3-10}
    & \texttt{BWT} $\Downarrow$ & 0.25 & 0.18 & 0.18 & 0.23 & 0.13 & 0.14 & 0.11 & \textbf{0.08} \\
    \midrule
    \multirow{2}{*}{$\mathcal{T}_{T_C}$\,--\,Curriculum} & \texttt{FWT} $\Uparrow$ & 0.12 & 0.14 & 0.13 & 0.14  & 0.16 & 0.17 & \textbf{0.2} & 0.19 \\
    \cmidrule(lr{1em}){3-10}
    & \texttt{BWT} $\Downarrow$ & 0.15 & 0.12 & 0.14 & 0.14 & 0.12 & 0.11 & 0.09 & \textbf{0.07} \\
    \midrule
    \multirow{2}{*}{$\mathcal{T}_{T_C}$\,--\,Random} & \texttt{FWT} $\Uparrow$ & 0.13 & 0.17 & 0.14 & 0.16  & 0.19 & 0.19 & \textbf{0.22} & 0.2 \\
    \cmidrule(lr{1em}){3-10}
    & \texttt{BWT} $\Downarrow$ & 0.23 & 0.16 & 0.21 & 0.21 & 0.13 & 0.13 & 0.1 & \textbf{0.08} \\

    \bottomrule
\end{tabular}
\label{tab:bwt-fwt}
\end{table*}
All the results reported in this section are averaged across 5 runs, each one involving a different random seed. We benchmark the performance of all the algorithms described in Section \ref{subsec:benchmark-algorithms} with respect to the metrics detailed in (\ref{eq:metrics}). We set the buffer size for all the replay-based methods to $200$. 
\subsection{Estimation Accuracy}
Figs.~\ref{fig:mse}(a)-(d) depict the performance in terms of average MSE at the end of all tasks while Figures~\ref{fig:mse}(e)-(h) depict the evolution of the MSE. First, we observe that the baseline NO-CL achieves the lowest MSE as expected which emphasizes the inevitable impact of catastrophic forgetting when continual learning is not taken into account. Besides, training on all tasks simultaneously (i.e., joint training method) leads to a lower bound on the MSE performance. Overall, the experience replay methods (i.e., ER, DER, DER++) outperform the other CL algorithms. It is also seen that they achieve the lowest MSE performance on both SNR and time coherence tasks. Moreover, we observe that the MSE values are lower in the curriculum order compared to the random one for both SNR and time coherence tasks. These results confirm the task ordering hypothesis as defined in Section \ref{subsec:task-protocol} which assumes that forgetting becomes less severe when the curriculum order is adopted.

\subsection{Task Transferability}
We present the results of the forward and backward transfer values in Table~\ref{tab:bwt-fwt}. It is observed that DER and DER++ achieve the best (i.e., highest) forward and (i.e., lowest) backward performance on all SNR and time coherence tasks with both curriculum and random task orders. Again,  the random task ordering harms both forward and backward transfers as compared to the curriculum ordering. Furthermore, it is interesting to note that the gap between random and curriculum orders for SNR tasks is smaller than the one for time coherence tasks. This is justified by the following two arguments. First, the ability to continuously learn is highly dependent on the number of total tasks. Hence, the fact that the number of time coherence tasks is higher than the number of SNR tasks makes learning with the former more challenging. Second, the SNR step size (2 dB) is more granular than the time coherence step size (5 symbol periods). This means that the obtained SNR tasks have a smoother increase in task difficulty (or equivalently a high correlation across tasks) as compared to the time coherence tasks, thereby improving the overall task transferability.

\section{Conclusion}
We have benchmarked the state-of-the-art continual learning methods on the Gauss-Markov Rayleigh fading channel estimation problem. After motivating how the SNR level and the time coherence lead to a significant shift in the distribution of the received power, we have varied these two parameters to define SNR and time coherence tasks. We have also examined the learning of these two tasks in both curriculum and random orders. The simulation results suggest that experience replay methods outperform the other continual learning approaches. This benchmark can be further extended to more diversified wireless communications tasks by varying the communication distance and/or the path-loss exponent, to determine whether or not the same experimental conclusions hold for them.

\bibliographystyle{IEEEtran}
\bibliography{IEEEabrv,references}

\end{document}